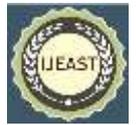

# SYNTHESIS AND OPTICAL CHARACTERIZATION OF PEROVSKITE LAYER FOR SOLAR CELL APPLICATION

Manoj Pandey, Dipendra Hamal
Department of Physics
School of Science
Kathmandu University, Dhulikel, Kavre, Nepal

Bijaya Basnet
Advanced Materials and Liquid Crystal Institute,
Kent State University, Kent, OH, 44242, USA

Bhim Kafle
Department of, Chemical Science & Engineering,
School of Engineering,
Kathmandu University, Nepal

*Abstract:* **Solvent engineering offers fine control over the photovoltaic efficiency, film morphology, and crystallization quality of perovskite films and also enables to optimize light transmittance and absorbance in solar cell applications. In the present work, the band gap and reflectance were reduced through solvent engineering. We found that perovskite thin films produced using DMF (Dimethyl form amide) solvent had a band gap that was 0.24 eV less than those produced using IPA (Isopropyl Alcohol) solvent. Perovskite thin films produced using DMF solvent also exhibited considerably lower solar spectrum reflectance.**

## I. INTRODUCTION:

The large absorption coefficient, flexibility, tunable band gap, high open-circuit voltage and higher power conversion efficiencies of the perovskite solar cell make it most suitable for the conversion of the solar energy into the electrical energy [1-6]. It is less expensive, combining the benefits of all-thin-film technology with high-efficiency solar energy harvesting [7]. $ABX_3$ represents the chemical formula of the perovskite material, where A could be organic cations, such as methyl ammonium ($MA^+$), form amidinium ($FA^+$) and cesium ($Cs^+$), B could be divalent metals, such as $Pb^{2+}$ and $Sn^{2+}$, and C could be an anion, such as $Cl^-$; $Br^-$; $I^-$, $SCN^-$ [1].The efficiency of perovskite solar cells has already surpassed 20%, therefore recent research has concentrated on band gap tuning, thermal stability, and minimizing moisture contact [1,8-9].

Developing the lead-free inorganic perovskite material, such as $Cs_2AgBiBr_6$, is of utmost important to address the toxicity and stability problems of conventional lead halide perovskite solar cells, despite the fact that $Cs_2AgBiBr_6$ film exhibits wide band, lower light absorption in the solar spectrum, and lower photo electronic conversion efficiency [10]. Zhang, Z. et al. [10] lowered the electronic bandgap from 2.18 eV to 1.64 eV using a hydrogenation technique that increased the photoelectric conversion efficiency to 6.4% with great environmental stability.

The organic-inorganic hybrid perovskite materials, such as methyl ammonium lead halide ($MAPbX_3$, X = Cl, Br, I) and form amidinium lead halide ($FAPbX_3$, X = Cl, Br, I), or their mixture, also exhibit humidity instability in in addition to being poor thermal stability since the organic components are easily vaporized [11-15]. The exceptional thermal stability of inorganic halide perovskites, like cesium lead halide, makes them potential candidates to address this issue. With the use of solvent-controlled development of the precursor film in a dry environment, Wang, P. et al. [11] improved the stability of the CsPbI3 film.

Jeon, N. et al. [16] employed the solvent engineering to enhance the efficiency of inorganic-organic hybrid perovskite solar cells, which also led to the growth of layers of perovskite that were incredibly homogenous and dense. With the aid of mixed solvent engineering, SungWon Cho et al. [17] strengthened the perovskite/ETL interface, decreased charge carrier recombination, and improved the morphology of the perovskite with reduced surface roughness. By employing the anti-solvent bath, Tian, Shi et al. [18] made a genuine attempt to develop a green solvent engineering technique and constructed perovskite films with a smoother surface and better performance.

As was already discussed, solvent engineering allows for precise control over the perovskite films' photovoltaic performance, film morphology, and crystallization quality. Additionally, it has been found that solvent engineering enables for balancing light transmittance and absorbance in solar cell applications [1]. As light absorbance exhibits a relationship with the band gap that is demonstrated in the Tauc plot [19-20], one can assume that the band gap could be tuned through solvent engineering. In the present work, we use two distinct solvents; DMF and IPA to generate the perovskite thin film. The band gap of perovskite thin films prepared with DMF solvent was found to be lower than that made with IPA solvent. We have investigated the impact of solvent engineering in enhancing the performance of perovskite solar cells by examining the optical characteristics of perovskite thin films.

II. METHOD:

**2.1 Preparation of methyl ammonium iodide (MAI) and methyl ammonium lead iodide (perovskite):**

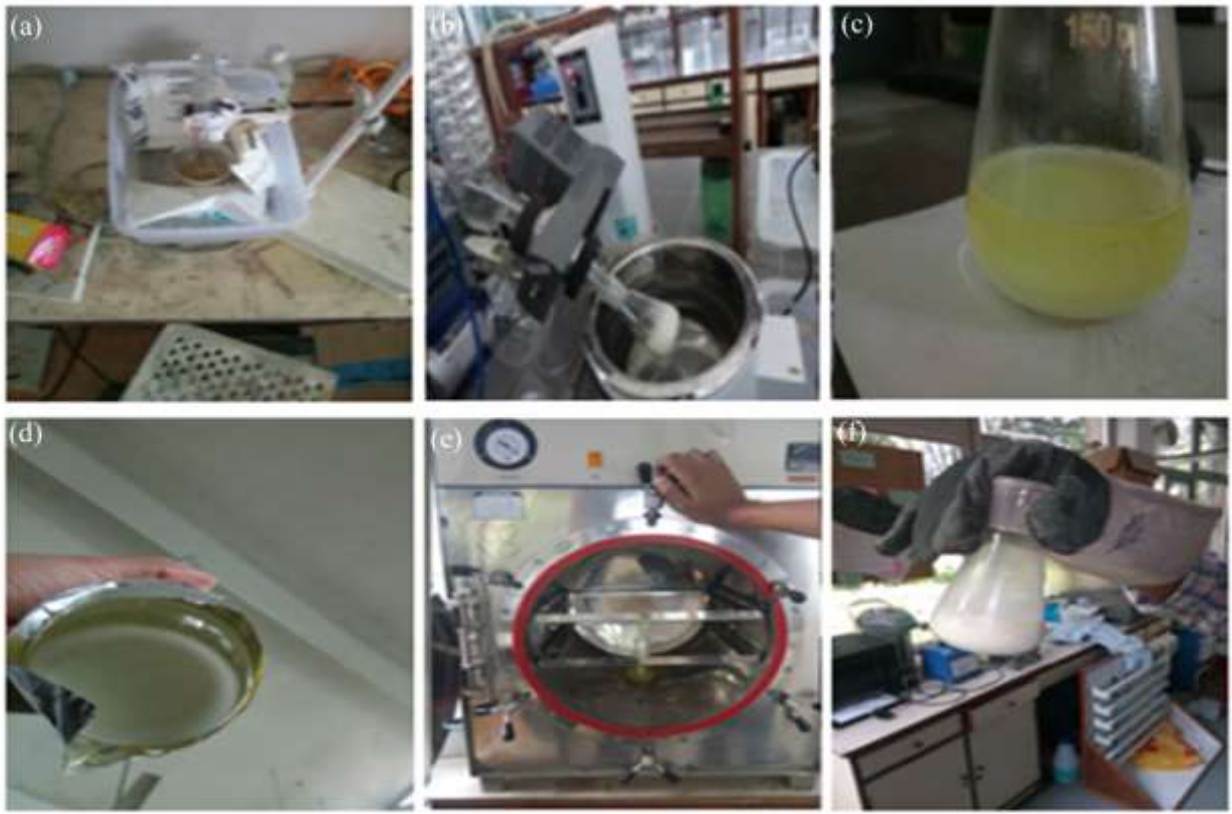

Figure 1:-Photographs during the preparation of methyl ammonium iodide, (a) mixture of methylamine and hydroiodic acid in ice bath, (b) mixing methylamine and hydroiodic acid solution in rotary evaporator , (c) MAI dissolved in ethanol, (d) MAI precipitated using diethyl ether, (e) MAI dried in vacuum oven, (f) pure MAI obtained from vacuum oven

In order to make methyl ammonium iodide (MAI), 27.8 mL of methylamine (33 wt.% in ethanol, Sigma-Aldrich) and 30 mL of hydroiodic acid (57 wt.% in water, Sigma-Aldrich) were combined drop wise and stirred at 0 °C for two hours. This solution's solvent was removed using a rotary evaporator operating at 50 °C, and a whitish powder (MAI) was recovered from the mixture. The MAI powder was then purified by being dissolved in 100% ethanol and precipitated by adding diethyl ether to the mixture. The MAI powder was then dried at 60 °C in a vacuum oven for an entire night [21]. The below chemistry schematic demonstrates the creation of methyl ammonium iodide:

$$CH_3NH_2 + HI \rightarrow CH_3NH_3I$$

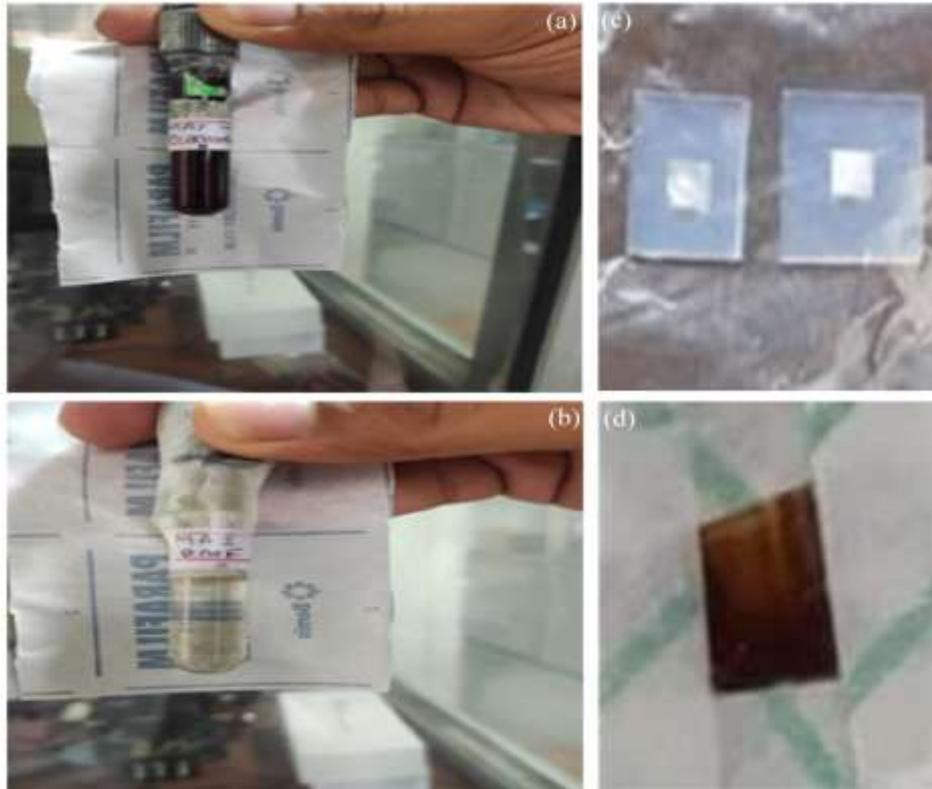

Figure 2:-Photographs of MAI solution in (a) iso-propyl alcohol, and (b) DMF; Photographs of (c) titanium dioxide films, and (d) Perovskite

Commercial compact titanium dioxide (C-TiO$_2$) is first applied on FTO coated glass using a spin coater at 1000 rpm for one minute, followed by two hours of annealing at 450°C. Then, using the Doctor blading process [22], mesoporous titanium dioxide (M-TiO$_2$) paste was deposited over the first layer of C-TiO$_2$ before being calcined at 450°C for two hours.

In DMF and Propan-2-0L (iso-propyl alcohol), the 0.55 M solution of MAI powder was produced separately at this time. Then, the spin coater is used to deposit both solutions on top of the mesoporous titanium dioxide film substrate for one minute at 1000 rpm, followed by an annealing step at 60°C to produce methyl ammonium lead iodide (perovskite) [23]. The optical properties of the perovskite films produced from MAI in DMF and MAI in iso-propyl alcohol are then reported. The following reaction illustrates how perovskite is formed:

$$CH_3NH_3I + PbI_2 \rightarrow CH_3NH_3PbI_3$$

**2.2 Optical characterization of the perovskite thin films**
Using a profilometer, we measured the perovskite thin films' reflectance and transmittance over the wavelength range of 200 to 800 nm. We used an integrating sphere and an Agilent Technologies Carry (Model: Carry 100 UV-Vis) for the absorbance measurement.

### III. RESULTS AND DISCUSSIONS

The perovskite thin film's UV-visible reflection (R) and transmission (T) spectra, as measured by the profilometer, are shown in figures 3 and 4 respectively. Over the glass substrate that has been coated with F-doped SnO$_2$ (FTO), compact titanium dioxide is deposited. The mesoporous titanium dioxide was coated over the compact titanium dioxide before the perovskite thin film was applied.

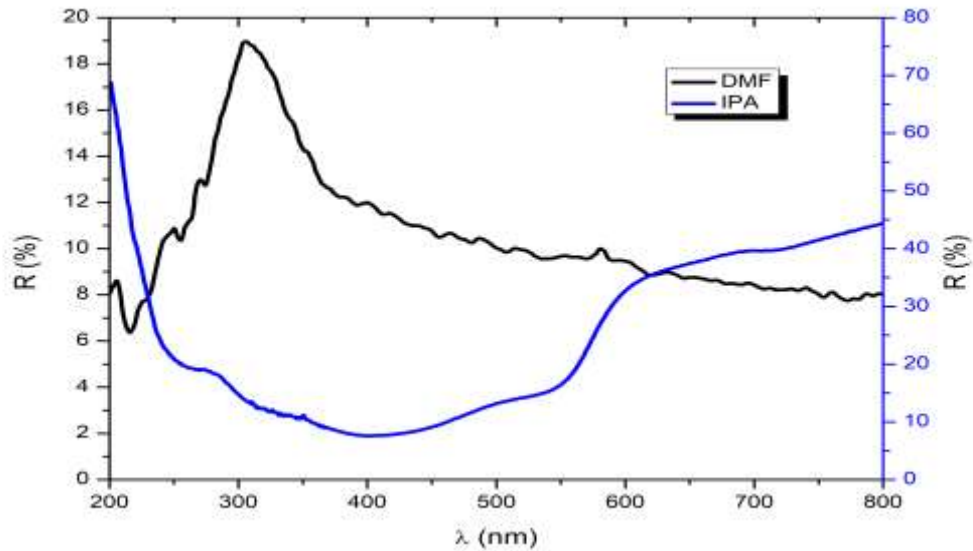

Figure 3: Reflectance vs wavelength of the perovskite thin films prepared by DMF and IPA solvents

It is noteworthy to mention that the mesoporous titanium dioxide, compact titanium dioxide, and F-doped $SnO_2$, all of which are transparent thin films with very little solar spectrum light absorption, while glass is the transparent thicker substrate with very low solar spectrum light absorption. Here, it is logical to assume that majority of the light intensity is absorbed in the perovskite layer.

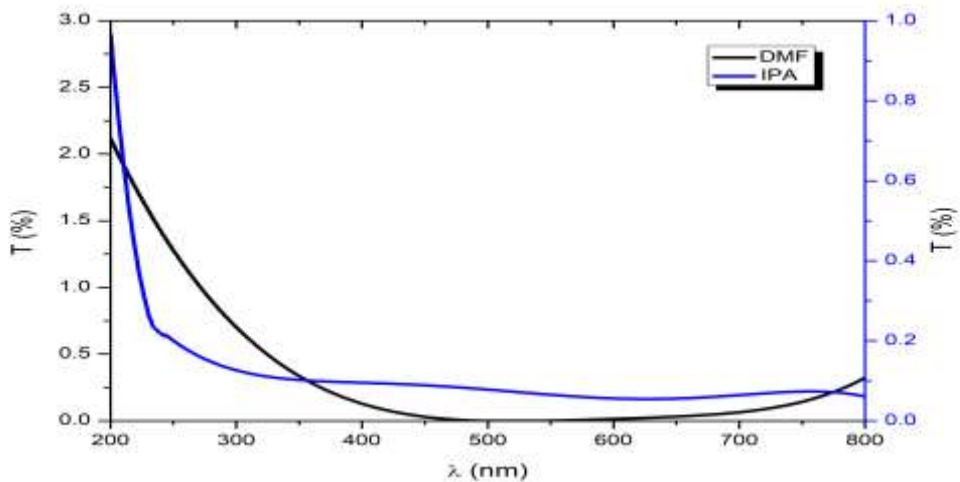

Figure 4: Transmittance vs wavelength of the perovskite thin films prepared by DMF and IPA solvents

Swanepoel, R[24] and Minkov, D. A. [25] devised the model to extract the optical properties of thin film deposited on the transparent substrate. These models, in turn, respectively use the interference pattern of the transmission and reflection spectra to calculate the thin film's thickness, refractive index, and absorption coefficient. In the reflection and transmission spectra, interference pattern vanishes when $x = \exp(-\alpha d) \to 0$, where α and d are the thin film's thickness and absorption coefficient, respectively [24-26]. Perovskite thin film exhibits higher absorption coefficient in the solar spectrum due to the relatively lower band gap of around 1.2-1.8eV [7, 27-28], which results $x \to 0$ and the disappearance of interference pattern in the reflection and transmission spectra. Also take into account that a lower absorption coefficient and a higher thickness might lead to $x \to 0$.

For the perovskite thin film with refractive index n and extinction coefficient k, one can find reflectance on normal incidence of light as [25-26],

$$R(\lambda) = \frac{(n-1)^2 + k^2}{(n+1)^2 + k^2} \quad [1]$$

Equation (1) is also known as Beer-Equation [29]. For the relatively smaller value of k (i.e. $k \ll n$), reflectivity is,

$$R(\lambda) = \frac{(n-1)^2}{(n+1)^2} \quad [2]$$

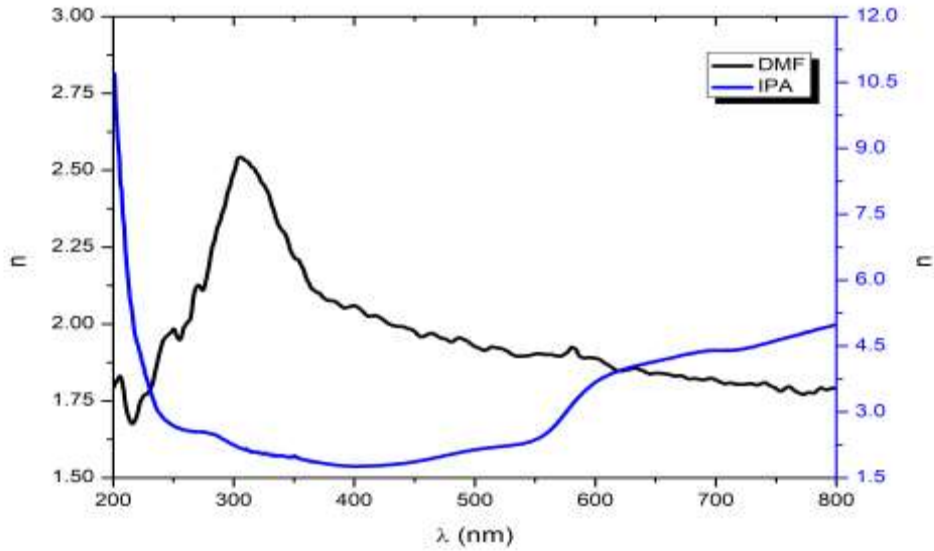

Figure 5: Refractive index vs wavelength of the perovskite thin films prepared by DMF and IPA solvents

By using the formula $n = \frac{1+\sqrt{R}}{1-\sqrt{R}}$ from Equation (2), we had calculated the refractive index from the reflection spectrum. Figure 5 displays the computed refractive index. In the absence of interference pattern, the transmission spectrum can be used to obtain the absorption coefficient in the region of strong absorption [24] as follows,

$$\alpha d = -\log(x) \approx -\log\left(\frac{(n-1)^3(n+s^2)}{16n^2 s} T\right) \quad [3]$$

where s is the refractive index of the substrate.
We use the Tauc plot [19-20] to calculate the optical band gap, which can be written as,

$$\alpha h\nu = A\left(h\nu - E_g\right)^r \quad [4]$$

where $h\nu$ is the energy of light in eV and $E_g$ is the band gap. Exponent r is 1/2 for perovskite thin film [27-28]. For the thin film with unknown thickness, we can plot $(\alpha h\nu d)^{1/r}$ vs $h\nu$, shown in Figure 3. As perovskite was coated over titanium dioxide, we use $s = 2.15$, which is the average value of refractive index in the visible and IR region [30-31]. The Tauc plot's extrapolated line intersects the x-axis at 1.51 eV and 1.75 eV respectively for DMF-perovskite and IPA-perovskite. The 1.5-1.6eV band gap for perovskite thin film was reported by Leal, Daniel Arturo Acuña, et al. in their study [28]. Hu, Mingyu, et al. [27] observed an even lower band gap value, particularly 1.4eV. The NBG perovskite absorber ($Cs_x$ ($FA_{0.83}$ $MA_{0.17}$)$_{(1-x)}Sn_{0.5}$ $Pb_{0.5}I_3$, $E_G \approx 1.26$ eV) and the WBG perovskite absorber ($FA_{0.8}$ $Cs_{0.2}$ ($I_{0.6}Br_{0.4}$)$_3$, $E_G \approx 1.78$ eV) exhibit considerably smaller and higher band gaps, respectively [7]. The range of band gap reported in the current work is in agreement with earlier research.

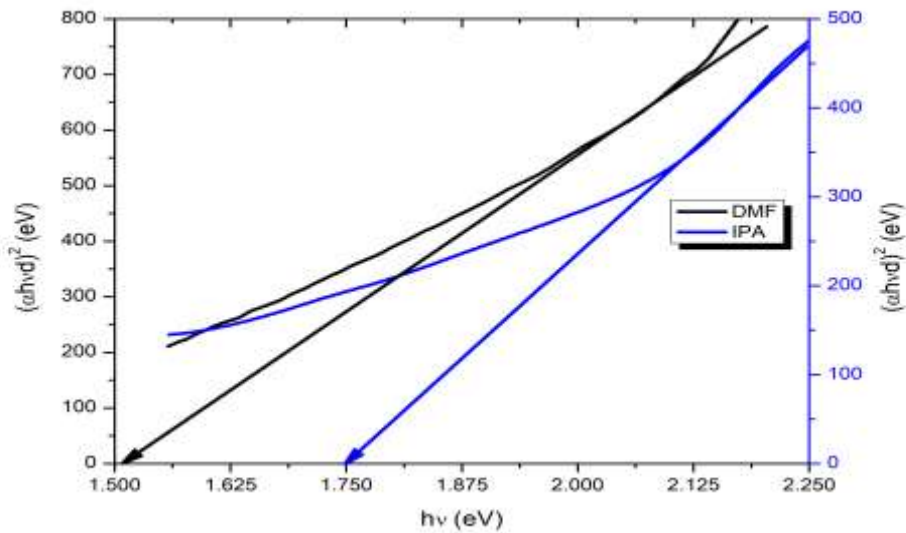

Figure 6: Tauc plot of the perovskite thin film films prepared by DMF and IPA solvents

From the beer-lambert law [32], the absorbance of the perovskite film can be written as,

$$A = \log_{10}\left(\frac{1}{T}\right) = 2 - \log_{10}(\%T) \quad [5]$$

Equation (5) was used to compute the absorbance of DMF-perovskite, which is depicted in Fig (6). In the visible light spectrum, DMF-perovskite exhibits extremely strong absorption. DMF-perovskite is hence highly suitable for use in solar cells. For the IPA-perovskite, the reflectance was found to be more than 40% in the visible and UV regions. Therefore, expression (5) can't precisely correlate the transmittance to the absorption coefficient. We utilized an Agilent Technologies Carry (Model: Carry 100 UV-Vis) with the help an integrating sphere for the absorbance measurement for IPA-perovskite thin film.

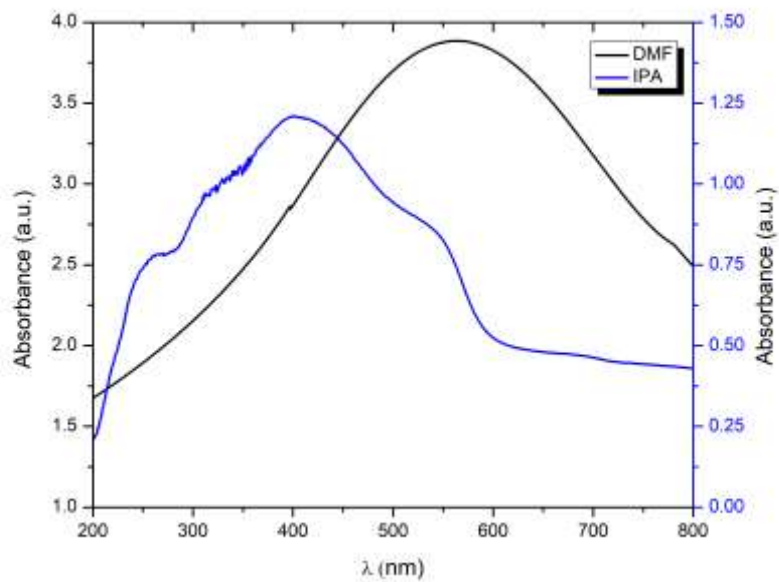

Figure 7: Absorbance vs wavelength of the perovskite thin films prepared by DMF and IPA solvents

We found that IPA perovskite thin film exhibits the stronger absorption in the UV and visible region than in IR region. The absorption peak was blue-shifted in comparison to the DMF perovskite thin film, which further supports the fact that the IPA solvent-produced perovskite thin film has a higher band gap than the DMF solvent-produced perovskite thin film, as photon energy is inversely proportional to wavelength. We found that the perovskite thin film produced using DMF solvent is more suited for use in high-efficiency solar cells due to its relatively lower band gap and optical absorption peak in the upper visible spectrum.

Zhu, Shijie, et al. [1] employed solvent engineering to raise the light absorbance and lower the light transmittance, and also optimized perovskite films in tandem devices. In the current work, we reduced the band gap and light reflectance through the solvent engineering.

## IV.  CONCLUSION:

The present study is mainly focused on the synthesis and characterization of perovskite film as aabsorber layer which is deposited over FTO and $TiO_2$ coated glass and to characterize its optical properties by varying the solvent of one of the precursor, MAI which is mixed with lead iodide to get perovskite film for its compatibility in solar cell. Perovskite thin film produced by using the DMF solvent exhibits the lower band gap and higher absorption in the solar spectrum in comparison to that produced using IPA as solvent, making it ideal for solar cell applications. In conclusion, solvent engineering enables the reduction of reflectance and band gap in perovskite thin films, which is important for the perovskite solar cells' efficiency gains.

Acknowledgements
This work was funded by University Grant Commission, Nepal (UGC Award No.: FRG-76/77-S&T-11).

Author contributions
Manoj Pandey and BhimKafle conceptualized the project. Manoj Pandey performed the experiments and analyzed the results. Bijaya Basnet analyzed the reflectance and transmittance spectra. Manoj Pandey and Bijaya Basnet wrote the manuscript. Dipendra Hamal conducted project administration. BhimKafle provided the laboratory resources.

Competing interests
The authors declare no competing interests.